\documentclass[twocolumn,showpacs,aps,floatfix,longtable,pre]{revtex4}

\usepackage{graphicx}
\usepackage{bm}

\begin{document}

\title{Ising antiferromagnet on the 2-uniform lattices}
\author{Unjong \surname{Yu}}
\affiliation{Department of Physics and Photon Science, Gwangju Institute of Science and Technology, Gwangju 61005, South Korea}
\email[E-mail me at: ]{uyu@gist.ac.kr}

\begin{abstract}
The antiferromagnetic Ising model is investigated on the twenty 2-uniform lattices
using the Monte-Carlo method based on the Wang-Landau algorithm and the Metropolis algorithm
to study the geometric frustration effect systematically.
Based on the specific heat, the residual entropy, and the Edwards-Anderson freezing
order parameter, the ground states of them were determined.
In addition to the long-range-ordered phase and the spin ice phase found in the
Archimedean lattices, two more phases were found.
The partial long-range order is long-range order with exceptional disordered
sites, which give extensive residual entropy.
In the partial spin ice phase, the partial freezing phenomenon appears:
Majority of sites are frozen without long-range order,
but the other sites are fluctuating even at zero temperature.
The spin liquid ground state was not found in the 2-uniform lattices.
\end{abstract}

\pacs{05.50.+q, 05.10.Ln, 64.60.De, 75.10.Hk}


\maketitle

\section{Introduction}

Recently, frustration has attracted intensive attention due to unexpected phenomena and exotic order
\cite{Balents10,Diep,Lacroix}.
The Ising model \cite{Ising25} always have a long-range-ordered ground state in two and higher dimension
without frustration \cite{Peierls36}.
With frustration, however, it has various ground states such as long-range-ordered
phase, spin glass, spin ice, and spin liquid phase \cite{Lacroix,Diep,Yu15}.
Frustration induced by geometric effect without disorder is called the geometric frustration \cite{Ramirez94}.
In this paper, the geometric frustration is investigated systematically
within the antiferromagnetic Ising model on two-dimensional 2-uniform lattices.

\begin{figure}[b]
\includegraphics[width=8.0cm]{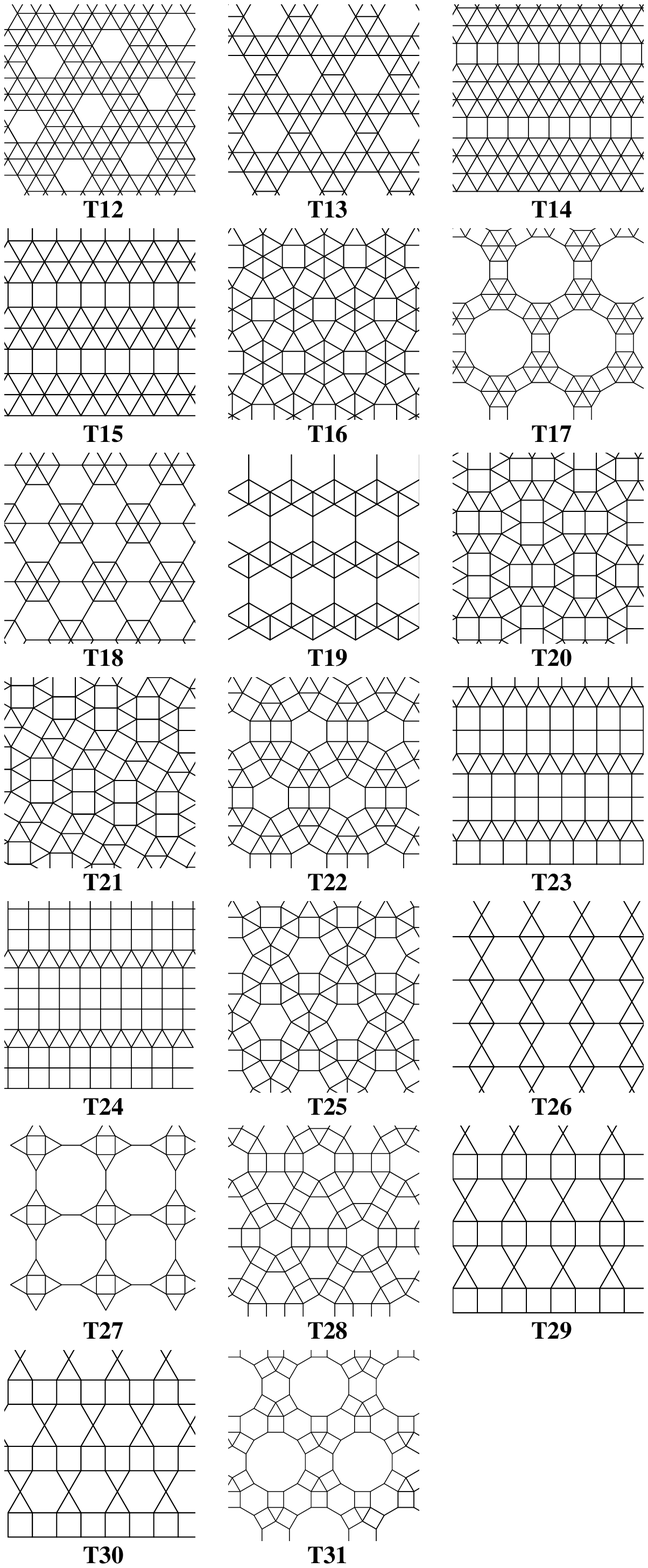}
\caption{The twenty 2-uniform lattices.}
\label{uni_fig}
\end{figure}

In a two-dimensional lattice made by regular polygons,
when there exist $k$ kinds of topologically equivalent vertices, it is called $k$-uniform lattice.
When there is only one kind (uniform tiling), it is also called the Archimedean lattice.
There are exactly eleven Archimedean lattices, and seven lattices among them are frustrated \cite{Grunbaum87}.
The ground state of the antiferromagnetic quantum Heisenberg model was investigated systematically:
the kagome and the star lattice were proposed as quantum spin liquid and
the other five lattices were classified tentatively as long-range-ordered phase \cite{Richter04,Farnell14}.
In the case of the antiferromagnetic Ising model, ground states of them are classified
into long-range-ordered phase (Shastry-Sutherland and trellis lattices),
spin ice (bounce, maple-leaf, and star lattices), and spin liquid (triangular and kagome lattices),
in the order of stronger frustration \cite{Yu15}.
Although the Archimedean lattice is a good starting point to study geometric frustration \cite{Krawczyk05},
seven lattices are not enough for a systematic study of various frustration effects,
and we present study of the frustration effect on the 2-uniform lattices in this work.
There exist exactly twenty 2-uniform lattices \cite{Grunbaum87}, as are listed in Fig.~\ref{uni_fig} and Table~\ref{uni_table}.
We labeled them from T12 to T31 after eleven Archimedean lattices.
Differently from the Archimedean lattices, which have corresponding natural material systems \cite{Richter04,Zheng14},
2-uniform lattices have not been found in nature, yet,
but it is expected to become possible to make artificial systems with a 2-uniform structure in the future.
At least, we can get an insight for the geometric frustration by systematic study on the 2-uniform lattices.

In this paper, we report detailed study of the antiferromagnetic Ising model on the 2-uniform lattices.
Specific heat, residual entropy, and freezing order-parameter are obtained to
identify the ground state. Finite temperature phase transitions
in weakly frustrated lattices are also investigated.

\begin{table*}[b]
\caption{\label{uni_table}Name, number of lattice points per basis ($B$), average coordination number ($z$),
antiferromagnetic ground state energy per bond ($E_g^{\rm AF}$), 
antiferromagnetic transition temperature ($T_{c}$),
exact residual entropy ($S_0$), and ground state for each 2-uniform lattice.
LRO means long-range order.
}
\begin{ruledtabular}
\begin{tabular}{cccccccc}
 & Name & $B$ & ~$z$~ & $E_g^{\rm AF}$ & $T_{c}^{\rm AF}$ & $S_0$ & Ground state\\
\hline
T21    & ($3^3, 4^2; 3^2, 4, 3, 4$)$_2$ &  8 & 5    & $-3/5$  &  1.193(1) & $\log(2)$ & LRO \\
T26    & ($3^2, 6^2; 3, 6, 3, 6$)       &  6 & 4    & $-2/3$  &  1.182(1) & $\log(2)$ & LRO \\
\hline
T13    & ($3^6; 3^4, 6$)$_2$            &  8 & 21/4 & $-3/7$  &  1.076(2) & $\log(2)(1+N/4)$ & Partial LRO \\
T17    & ($3^6; 3^2, 4, 12$)            & 14 & 30/7 & $-3/5$  &  1.210(1) & $\log(2)(1+N/7)$ & Partial LRO \\
T18    & ($3^6; 3^2, 6^2$)              &  7 & 30/7 & $-3/5$  &  1.324(1) & $\log(2)(1+N/7)$ & Partial LRO \\
\hline
T14    & ($3^6; 3^3, 4^2$)$_1$          &  4 & 11/2 & $-5/11$ &  0.17(1)  & $\log(8)L$ & 1-direction LRO \\
T15    & ($3^6; 3^3, 4^2$)$_2$          &  3 & 16/3 & $-1/2$  &  0.27(1)  & $\log(4)L$ & 1-direction LRO \\
T19    & ($3^4, 6; 3^2, 6^2$)           &  4 & 9/2  & $-5/9$  &  0        & $\log(2)L$ & 1-direction LRO \\
T23    & ($3^3, 4^2; 4^4$)$_1$          &  3 & 14/3 & $-5/7$  &  0        & $\log(2)L$ & 1-direction LRO \\
T24    & ($3^3, 4^2; 4^4$)$_2$          &  4 & 9/2  & $-7/9$  &  0        & $\log(2)L$ & 1-direction LRO \\
\hline
T30    & ($3, 4^2, 6; 3, 6, 3, 6$)$_2$  &  5 & 4    & $-3/5$  &  0        & $\log(2)(L+N/5)$ & Partial 1-direction LRO \\
\hline
T20    & ($3^3, 4^2; 3^2, 4, 3, 4$)$_1$ & 12 & 5    & $-7/15$ & -         & $\left[0.18472(1)\right] N$          & Spin ice\\
T22    & ($3^3, 4^2; 3, 4, 6, 4$)       & 12 & 9/2  & $-5/9$  & -         & $\left[0.1258(2)\right] N$           & Spin ice \\
T25    & ($3^2, 4, 3, 4; 3, 4, 6, 4$)   & 12 & 9/2  & $-5/9$  & -         & $\left[0.083(3)\right] N$            & Spin ice \\
T28    & ($3, 4^2, 6; 3, 4, 6, 4$)      & 18 & 4    & $-2/3$  & -         & $\left[0.08401(1)\right] N$          & Spin ice \\
\hline
T12    & ($3^6; 3^4, 6$)$_1$            & 12 & 11/2 & $-13/33$& -         & $\left[0.172(1)\right] N$            & Partial spin ice \\
T16    & ($3^6; 3^2, 4, 3, 4$)          &  7 & 36/7 & $-4/9$  & -         & $\left[(\log(2)+0.32306595)/7\right] N$ & Partial spin ice \\
T27    & ($3, 4, 3, 12; 3, 12^2$)       &  8 & 7/2  & $-3/7$  & -         & $\left[0.37757(1)\right] N$          & Partial spin ice \\
T29    & ($3, 4^2, 6; 3, 6, 3, 6$)$_1$  &  5 & 4    & $-3/5$  & -         & $\left[0.18376(2)\right] N$          & Partial spin ice \\
T31    & ($3, 4, 6, 4; 4, 6, 12$)$_2$   & 18 & 10/3 & $-11/15$& -         & $\left[0.12885(2)\right] N$          & Partial spin ice \\
\end{tabular}
\end{ruledtabular}
\end{table*}

\section{Model and method}
The Ising model Hamiltonian in this work is as follows.
\begin{eqnarray}
H = -J \sum_{\langle i,j \rangle} S_{i} S_{j} ~.
\label{Eq:Ising}
\end{eqnarray}
The spin at the $i$-th site $S_{i}$ may take the values of $+1$ or $-1$, only.
The summation $\langle i,j \rangle$ runs for all the nearest-neighbor pairs, excluding double counting.
The coupling constant $J$ and the Boltzmann constant $k_B$ are set to be $J=-1$ and $k_B=1$.
Only antiferromagnetic interaction is considered within this paper.

The calculation was done for parallelepiped lattices with number of sites $N=B\times L\times L$,
where $B$ and $L$ are the number of sites per unit cell and the linear size, respectively. 
The periodic boundary condition is used.
To simulate frustrated systems, we used 
the Wang-Landau algorithm \cite{Wang01}.
The Wang-Landau algorithm calculates the density of states $\rho(E)$ (DOS)
as a function of energy $E$ directly by the random walk
with the transition probability $P(i\rightarrow j) = \min \{1 , \rho(E_i) / \rho(E_j)\}$.
At each step, the histogram $h(E_i)$ and
the DOS $\rho(E_i)$ of the present energy $E_i$ is adjusted
by $h(E_i)\rightarrow h(E_i) + 1$ and $\rho(E_i)\rightarrow f_n \, \rho(E_i)$
with an amplification factor $f_n > 1$ until the histogram $h(E_i)$ becomes flat enough.
When the standard deviation of the histogram is less than 4\% of its average,
the normalization of $\sum_{i} \rho(E_i) = 2^{N}$ is performed and
a new scan starts with an empty histogram [$h(E_i)=0$] and a smaller value of $f_{n+1} = \sqrt{f_n}$.
The whole simulation begins with an initial amplifier $f_0 = e$ and stops when $f_n$ approaches enough to 1:
 $f_n < \exp(10^{-10})$.
If the simulation is successful, the error is expected to be the same order as $\log(f_n)$.
Although the Wang-Landau method is slow and can be applied to only small clusters,
it gives reliable results for the frustrated systems because it is not stuck to a metastable state.
After the DOS $\rho(E)$ is obtained, the specific heat $c(T)$ can be calculated easily as a function of temperature $T$:
\begin{eqnarray}
c(T) = \frac{1}{T^2} \left\{ \langle E^2 \rangle _T - \left( \langle E \rangle _T \right)^2 \right\} \\
\mbox{with } ~ \langle X \rangle _T =  \frac{\sum_{i} X_i \, \rho(E_i) \, e^{-E_i/T}}{\sum_{i} \rho(E_i) \, e^{-E_i/T}} ~.
\label{Wang_Landau}
\end{eqnarray}
The residual entropy $S_0$ can be calculated also directly from the DOS ($\rho(E)$) obtained
by the Wang-Landau algorithm: $S_0 = \log[\rho(E_0)]$, where $E_0$ is the lowest energy.
Since the Wang-Landau algorithm can give only static information,
we also used the Metropolis algorithm \cite{Metropolis53} to study fluctuation around
     a ground state.

\section{Results and discussion}


\begin{figure}
\includegraphics[width=8.4cm]{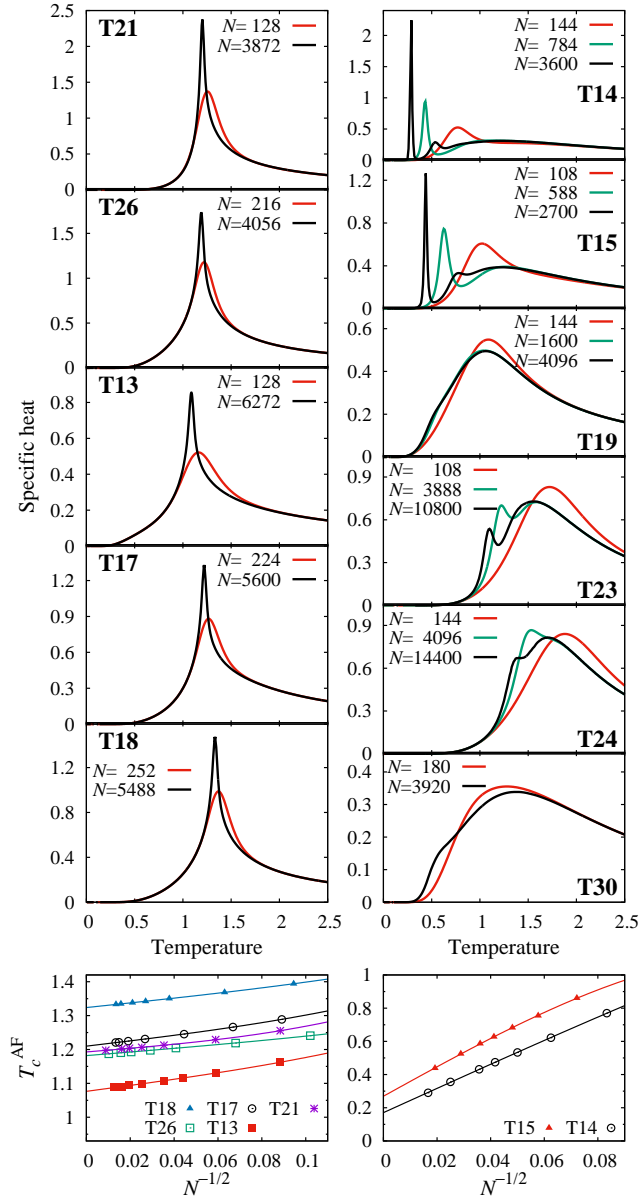}
\caption{(Color online) Specific heat as a function of temperature (upper panels) and
finite size-scaling to find the critical temperature (lower panels) in the eleven weakly frustrated 2-uniform lattices,
which have long-range-ordered ground states.
Left and right panels are for lattices with long-range ordering in two and one directions, respectively.
Solid lines in the bottom panels are results of fitting for the 
antiferromagnetic critical temperatures ($T_c^{\mathrm{AF}}$) 
by $T_c^{\mathrm{AF}}(N) = T_c^{\mathrm{AF}}(\infty)+\lambda N^{-\nu/2}(1+b N^{-\omega/2})$
with $\nu=1$ and $\omega=2$. The temperature is in the unit of $|J|/k_B$.
}
\label{Cv1}
\end{figure}

\begin{figure}
\includegraphics[width=8.2cm]{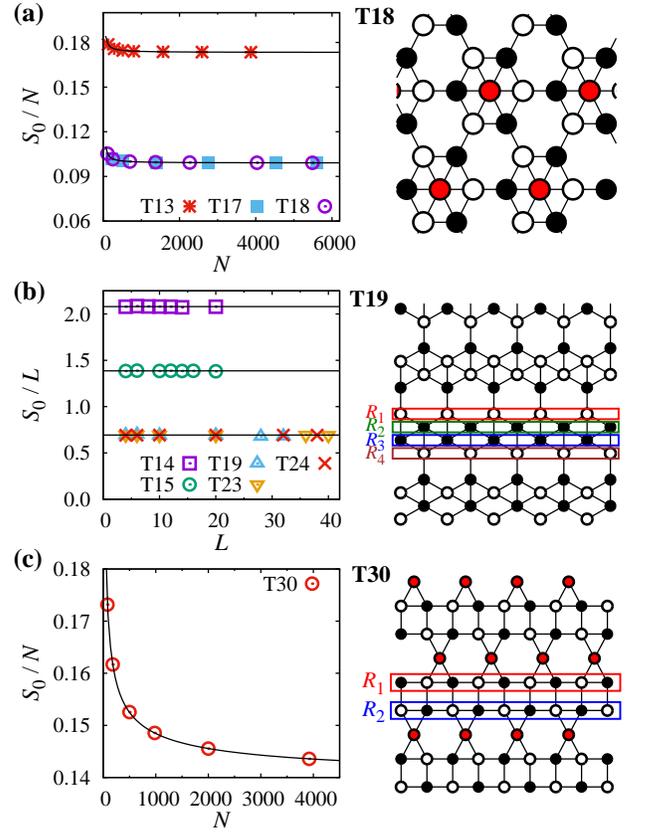}
\caption{(Color online) Residual entropy $S_0$ calculated by the Wang-Landau method
for weakly frustrated lattices with long-range-ordered ground state.
Lines represent the exact results for each lattice:
$S_0=\log(2)(1+N/4)$ and $S_0=\log(2)(1+N/7)$ in (a),
$S_0=\log(8)L$, $S_0=\log(4)L$, and $S_0=\log(2)L$ in (b),
and $S_0=\log(2)(L+N/5)$ in (c).
Examples of the ground states for T18, T19, and T30 are shown in the right panels.
Black and white circles represent up and down spin sites, respectively.
Red circles are for disordered sites with random spin.}
\label{entropy1}
\end{figure}

From the specific heat (Fig.~\ref{Cv1}) and residual entropy data (Fig.~\ref{entropy1})
obtained by the Wang-Landau method,
eleven lattices were classified to have long-range-ordered ground states.
Two of them (T21 and T26) have conventional long-range order with residual entropy $S_0 = \log(2)$,
which was confirmed by the Wang-Landau calculation.
The degeneracy of two is from the Kramers degeneracy.
Other three lattices (T13, T17, and T18) have partial long-range order,
which means the long-range order except part 
of the lattice that keeps disordered even at zero temperature.
Therefore, the residual entropy is extensive like the spin ice and the spin liquid.
For example, a ground state of lattice T18 is shown in the right panel of Fig.~\ref{entropy1}(a).
Without sites in the center of hexagons, which are represented by red circles,
there is no frustration and the ground state
is uniquely determined except the Kramers degeneracy. Since the site can have any spin state between up and down,
the degeneracy is $W_0=2\times2^{L^2}=2\times2^{N/7}$ and the residual entropy becomes extensive:
$S_0 = \log(W_0) = \log(2)(1+N/7)$. In the same way, the residual entropy of T13 and T17 can be understood, too.
In the case of T13, two spins out of eight spins in a unit cell are disordered and $S_0 = \log(2)(1+N/4)$.
These behaviors are well confirmed by the Wang-Landau calculations.
The partial long-range order has been observed in a number of models with frustration.
The classical Heisenberg model with uniaxial exchange anisotropy on the triangular lattice
has a partial order, where 2/3 spins make a long-range-ordered honeycomb lattice
while the other 1/3 spins are disordered \cite{Nagai93}. The body-centered-cubic (bcc) lattice may have the partial order
with nearest-neighbor (NN) and next-nearest-neighbor (NNN) interactions, too \cite{Santamaria97}.
The Potts model shows a partial order in a few Laves lattices (diced, union jack, and centered diced lattices)
\cite{Kotecky08, Chen11, Qin14}. As for the Ising model, it was found in the
fully-frustrated simple-cubic lattice \cite{Blankschtein84,Diep85},
the stacked triangular lattice \cite{Blankschtein84b},
the kagome and the bcc lattices with NN and NNN interactions \cite{Azaria87,Azaria89},
and the anisotropic centered honeycomb lattice \cite{Diep91}.
The partial order was also proposed in the quantum Heisenberg model \cite{Quartu97,Santamaria97}
and the periodic Anderson model \cite{Hayami11}.
Partial order phenomena are also observed experimentally in a few antiferromagnetic materials
such as GdInCu$_4$ \cite{Nakamura99}, Gd$_{2}$Ti$_{2}$O$_{7}$ \cite{Stewart04}, and Sr$_2$YRuO$_6$ \cite{Granado13}.


The five lattices have finite temperature order-disorder transition indicated by diverging
specific heat. The temperature of maximum specific heat for various cluster size
follows the scaling behavior of the two-dimensional Ising universality class \cite{Pelissetto02}:
\begin{eqnarray}
T_c^{\mathrm{AF}}(N) = T_c^{\mathrm{AF}}(\infty)
       +\frac{\lambda}{N^{\nu/2}} \left(1+\frac{b}{N^{\omega/2}}\right)
\label{fss}
\end{eqnarray}
with $\nu=1$ and $\omega=2$. The two variables $\lambda$ and $b$ are lattice-dependent fitting parameters.
By this method, the critical temperatures were obtained as shown in Fig.~\ref{Cv1} and Table~\ref{uni_table}.

The other six lattices (T14, T15, T19, T23, T24, and T30) have the long-range order
only in one direction like the Trellis lattice \cite{Yu15}.
For example, T19 is made by stripes, each of which is composed of four rows.
(See right panel of Fig.~\ref{entropy1}(b).) In the ground state, all the rows are ordered
with the same spin in the horizontal direction.
The spins of the first ($R_1$) and the second ($R_2$) rows are uniquely determined by
the spin of the last row of its adjacent stripe. However, the third ($R_3$) and  the fourth ($R_4$) rows
may choose one between up-down or down-up spins. In other words, when the spin of $R_1$ is fixed to be
up, there are two degenerate spin configurations: up-down-up-down and up-down-down-up
 for $R_1$, $R_2$, $R_3$, and $R_4$. Since each stripe has the degeneracy of two, the 
residual entropy becomes $S_0 = \log\left(2^L\right) = \log(2) L$.
By the same reason, the residual entropies of T23 and T24 are also $S_0 = \log(2) L$.
T14 and T15 have degeneracy of $2^3$ and $2^2$ per unit stripe and their residual entropies
are $S_0 = \log(8) L$ and $S_0 = \log(4) L$, respectively.
As shown in Fig.~\ref{entropy1}(c), T30 has a partial long-range order in one direction.
When $R_1$ is ordered as $\uparrow\downarrow\uparrow\downarrow\uparrow\downarrow\cdots$,
adjacent $R_2$ should be ordered as $\downarrow\uparrow\downarrow\uparrow\downarrow\uparrow\cdots$.
Thus, each stripe composed of $R_1$ and $R_2$ has a degeneracy of two. Sites between the two stripes
(red circles in Fig.~\ref{entropy1}(c)) may have any spin and they have degeneracy of two per site.
Therefore, the residual entropy of T30 becomes $S_0 = \log(2)(L+N/5)$.

As shown in Fig.~\ref{Cv1}, specific heat data of T14 and T15 behave similar to the Trellis lattice. 
They show size-independent broad peak and size-dependent sharp diverging peak at lower temperature,
where the one-direction ordering occurs. The temperature of maximum specific heat fits well with Eq.~(\ref{fss})
to give the critical temperature of $T_c^{\mathrm{AF}}=0.17(1)$ and $0.27(1)$ for T14 and T15, respectively.
Differently from the trellis lattice, the critical exponents $\nu$ and $\omega$ are the same
as the two-dimensional Ising model within statistical error.
As for T19, T23, T24, and T30 no finite temperature phase transition was found.


\begin{figure}[b]
\includegraphics[width=8.2cm]{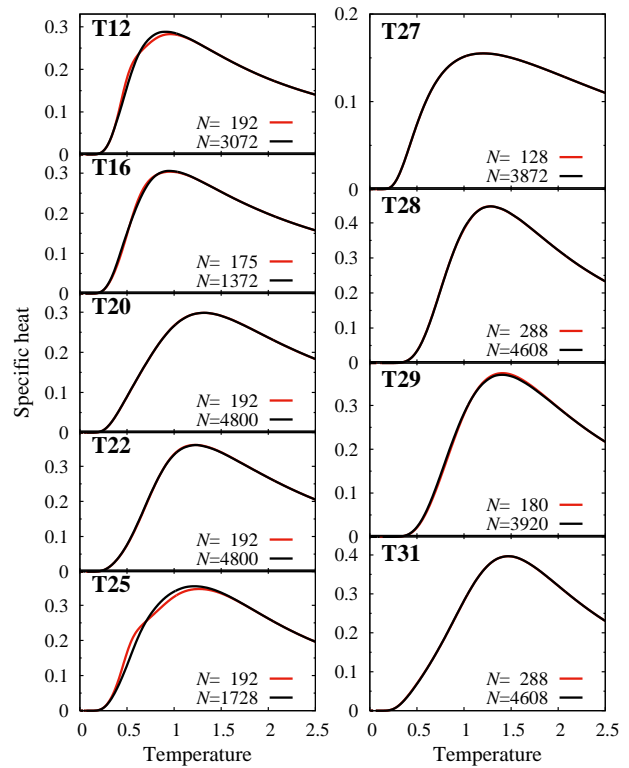}
\caption{(Color online) Specific heat as a function of temperature for the nine strongly frustrated 2-uniform lattices. 
}
\label{Cv2}
\end{figure}

\begin{figure}[b]
\includegraphics[width=8.2cm]{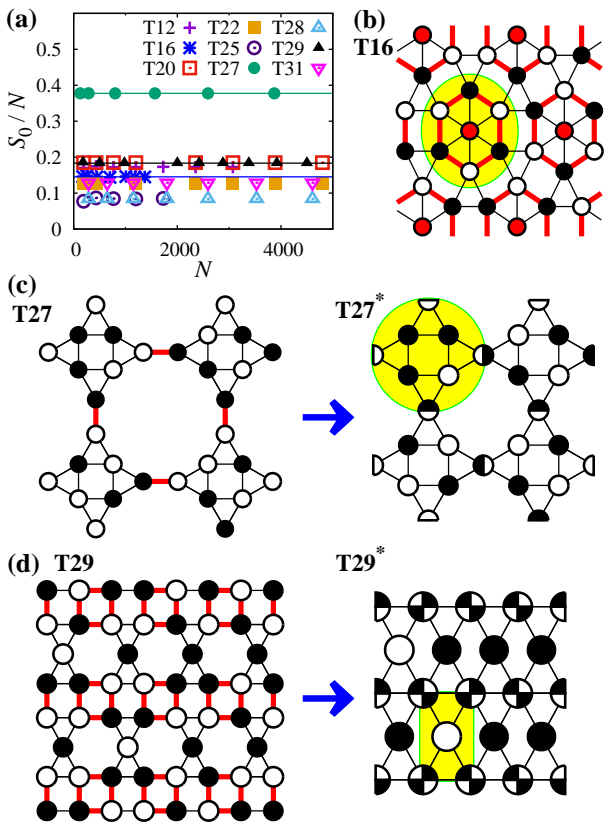}
\caption{(Color online) (a) Residual entropy for the 9 strongly frustrated 2-uniform lattices.
Horizontal solid lines are $S_0/N = \log(20.5)/8$, $\log(2.5)/5$, and $(0.32306595+\log(2))/7$
for T27, T29, and T16, respectively. (b) A ground state for T16. Black, white, and red circles
represent sites with up-spin, down-spin, and disordered sites. Thick red lines indicate unfrustrated
links. (c) and (d) show examples of ground states for T27 and T29.
By merging the unfrustrated links (thick red lines), they can be transformed into T27$^*$ and T29$^*$,
which have the same residual entropy per unit cell as T27 and T29, respectively.
Yellow ellipses in (b), (c), and (d) show unit cells. 
}
\label{ent2}
\end{figure}

The other nine lattices were found to be disordered even at zero temperature.
Their specific heat data as a function of temperature are shown in Fig.~\ref{Cv2}.
They do not have a diverging peak and the size-dependence is very small, which
implies absence of long-range correlation. Clearer evidence is
the residual entropy, which 
is proportional to the number of spins within the cluster.
They are plotted in Fig.~\ref{ent2} and listed in Table~\ref{uni_table}.
Some of them can be calculated analytically. A unit cell of T16 includes
seven spins, which makes a hexagon with an extra central spin. (See Fig.~\ref{ent2}(b).)
Without a central spin (red circle), there is no frustration and the
six spins have spin configuration of
($\uparrow\downarrow\uparrow\downarrow\uparrow\downarrow$)
or ($\downarrow\uparrow\downarrow\uparrow\downarrow\uparrow$).
Neighboring hexagons prefer to have different spin configuration with each other, but
it is not always possible because they form a triangular lattice.
Thus, the residual entropy by hexagons per unit cell is the same as the 
triangular lattice (T1). Since the central spin may have any direction without
changing the energy, it gives extra residual entropy to the lattice.
Therefore, the residual entropy is $S_0^{\rm T16}/N = [S_0^{\rm T1}/N+\log(2)]/7$.
The residual entropy of the triangular lattice $S_0^{\rm T1}$ is exactly known:
$S_0^{\rm T1}/N = (3/\pi) \int_{0}^{\pi/6} \log(2 \cos w) dw \approx 0.32307$ \cite{Wannier50}.
The residual entropy of T27 and T29 can be understood by the Pauling's method applied to the ice \cite{Pauling35}.
This method gives approximate but very accurate estimation in
corner-sharing frustrated systems \cite{Yu15}. For example, the residual entropy of
the kagome lattice (corner-sharing triangles) can be estimated
by $S_0^{\rm T8} \approx \log\left[2^N (6/8)^{2N/3}\right] \approx 0.50136 N$ \cite{Liebmann},
where $(6/8)$ means six states out of $2^3$ states give minimum energy within a triangle
and $2N/3$ means number of triangles in the lattice.
This value is smaller than the exact result by 0.1\% \cite{Kano53}.
In the case of T27, a unit cell, which is star-shaped, is connected with neighboring
unit cells with unfrustrated links (thick red lines in Fig.~\ref{ent2}(c)).
When the two spins connected by unfrustrated links are merged,
T27 is transformed into T27$^*$, which has the
same residual entropy per unit cell as T27. The residual entropy of T27$^*$ is estimated
by $S_0^{\rm T27^*} \approx \log\left[2^N (82/2^8)^{N/6}\right] \approx 0.50340 N$.
$(82/2^8)$ means only 82 states out of $2^8$ states satisfy the minimum
energy condition in a star-shaped cluster composed of eight spins.
Since T27 and T27$^*$ have the same residual entropy per unit cell,
$S_0^{\rm T27} = (6/8) S_0^{\rm T27^*} \approx 0.37755$.
The same method can be applied to T29. By merging four spins connected
by square-shaped unfrustrated links, T29 is transformed
into T29$^*$. (See Fig.~\ref{ent2}(d).) The residual entropy of T29 is estimated
by $S_0^{\rm T29} = (2/5) S_0^{\rm T29^*} \approx (2/5) \log\left[2^N (10/2^4)^{N/2}\right] \approx 0.18326$.
These estimations are smaller than the numerical results by the Wang-Landau algorithm
by 0.005\% and 0.3\% for T27 and T29, respectively.
The other six lattices (T12, T20, T22, T25, T28, and T31) are too complicated to get
analytic results or Pauling's estimates. They are in-between 0.08 and 0.18 per spin,
which is about half of that of the triangular lattice.
This is because there are some unfrustrated links, which reduce the
residual entropy of the lattice relative to the triangular lattice.
Thick red links in Fig.~\ref{ent2}(b)-(d) are examples.

\begin{figure}[b]
\includegraphics[width=8.2cm]{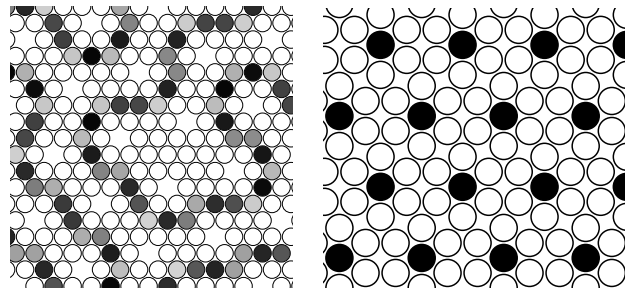}
\caption{Freezing order parameter $q_{\mathrm{EA},i}$ at $T=0.0001$
for T12 (left) and T16 (right) in the partial spin ice phase.
White, gray, and black circles represent
frozen sites ($q_{\mathrm{EA},i}=1$), slowly fluctuating sites, and quickly fluctuating sites
($q_{\mathrm{EA},i}=0$), respectively.
}
\label{partial_SI}
\end{figure}

Another important property to study for frustrated systems is the freezing. Spins are frozen in the
spin glass phase and the spin ice phase, as well as in the long-range-ordered phase.
When spins are not frozen and fluctuate even at zero temperature, it is called the spin liquid.
Since DOS does not give information about freezing phenomena, we performed another simulation using
the single-site update of the Metropolis algorithm \cite{Metropolis53}.
The degree of freezing can be measured by the Edwards-Anderson order parameter $q_{\mathrm{EA}}$ \cite{EA},
which was proposed to study freezing phenomena in spin glass systems. It can be used generally to study
freezing phenomena including the spin ice. We adopted the implementation of Ref.~\onlinecite{Ngo14}:
\begin{eqnarray}
q_{\mathrm{EA},i} = \frac{1}{M} \left| \sum_{t}^{M} S_i(t) \right| , \\
q_{\mathrm{EA}} = \frac{1}{N} \sum_{i}^{N} q_{\mathrm{EA},i} ~.\label{freezing_Eq} 
\end{eqnarray}
$M$ is the number of Monte-Carlo steps for measurement after equilibration,
which is fixed to be $M=2\times10^6$ in this work.
The spin to be tested for flipping is chosen randomly, and each spin is tested once per one Monte-Carlo step on average.
At high temperature, spins fluctuate fast and $q_{\mathrm{EA}} \approx 0$.
Except the spin liquid, it begins to increase at the transition temperature $T_c$ or freezing temperature $T_f$
as temperature is lowered.
For lattices with long-range-ordered ground state (T21, T26, T14, T15, T19, T23, and T24)
and spin ice ground state (T20, T22, T25, and T28), all the spins are frozen and 
$q_{\mathrm{EA}}$ approaches 1 at zero temperature.
For lattices with partially-long-range-ordered ground state (T13, T17, T18, and T30),
spins are frozen except disordered spins and $q_{\mathrm{EA}}$ approaches a value in-between zero and one.
For example, $q_{\mathrm{EA}}$ approaches $6/7$ in T18 as temperature goes to zero.
No spin liquid ground state was found among the 2-uniform lattices.

Finally we found a new phase, partial spin ice, which appears in T12, T16, T27, T29, and T31,
where majority spins are frozen at zero temperature without long-range order,
but the other spins are fluctuating. In Fig.~\ref{partial_SI},
$q_{\mathrm{EA},i}$ is shown for T12 and T16. Disordered spins can be connected to make a string
like T12 or isolated like T16. Since frozen region is larger and spans the whole lattice,
it would be natural to consider them as a spin ice with exceptional fluctuating parts, and
we named them as the partial spin ice.

\begin{figure}[b]
\includegraphics[width=8.2cm]{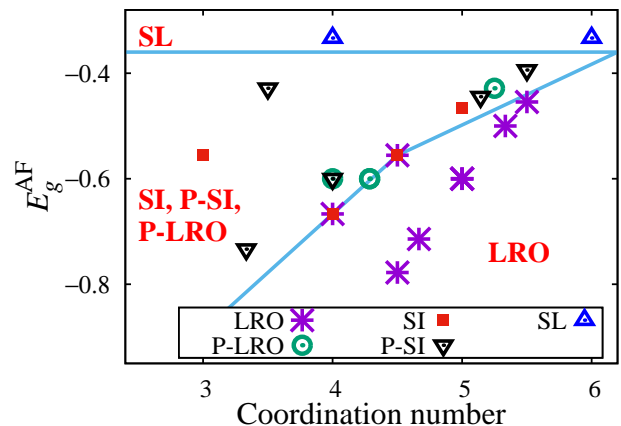}
\caption{(Color online) Ground states of the Ising antiferromagnet on twenty seven frustrated Archimedean and 2-uniform lattices
represented with their ground state energy per bond $E_g^{\mathrm{AF}}$ and
the coordination number $z$. Lower-right corner and upper region indicate weak and strong frustration, respectively.
LRO, P-LRO, SI, P-SI, and SL means long-range order, partial long-range order, spin ice, partial spin ice,
and spin liquid, respectively.
}
\label{phase}
\end{figure}

The ground states of seven frustrated Archimedean lattices and twenty 2-uniform lattices are
shown with their ground state energy per bond $E_g^{\mathrm{AF}}$ and
the coordination number $z$ in Fig.~\ref{phase}. Higher $E_g^{\mathrm{AF}}$ and smaller $z$
indicates stronger frustration and tends to induce spin liquid ground state \cite{Kobe76,Kobe95,Richter04}.
To the contrary, lattices with lower $E_g^{\mathrm{AF}}$ and larger $z$ tend to
have a long-range-ordered ground state.
In-between, spin ice, partial spin ice, and partial long-range-ordered ground states exist.
The general trend is confirmed as shown in Fig.~\ref{phase},
but we found it is not an absolute criterion. T7 (bounce lattice), T26, and T28
have the same coordination number and ground state energy per bond,
but T7 and T28 have spin ice ground state while T26 has
long-range-ordered ground state.

\section{Summary}

We studied systematically the geometric frustration effect of the Ising antiferromagnet
on the 2-uniform lattices. From the results of specific heat, residual entropy, and
freezing order parameter, we classified ground states of them into 
long-range order, partial long-range order, spin ice, and partial spin ice.
In comparison with the Archimedean lattices \cite{Yu15}, partial long-range order and 
partial spin ice phase were additionally found but spin liquid phase is missing.
Table~\ref{uni_table} summarizes the results.

\section*{Acknowledgments}
This work was supported by the GIST Research Institute (GRI) in 2016.

\bibliography{uni}

\end{document}